\documentclass[10pt,letterpaper]{article}
\usepackage[top=0.85in,left=2.75in,footskip=0.75in]{geometry}

\usepackage{amsmath,amssymb}
\usepackage[utf8x]{inputenc}
\usepackage{cite}
\usepackage{nameref}
\usepackage[colorlinks=true,allcolors=blue]{hyperref}
\usepackage{microtype}
\DisableLigatures[f]{encoding = *, family = * }
\usepackage{array}
\usepackage{xcolor}
\usepackage{ulem}

\raggedright
\usepackage[aboveskip=1pt,labelfont=bf,labelsep=period,justification=raggedright,singlelinecheck=off]{caption}

\makeatletter
\renewcommand{\@biblabel}[1]{\quad#1.}
\makeatother

\date{}

\usepackage{lastpage,fancyhdr,graphicx}
\usepackage{epstopdf}
\fancyheadoffset[L]{2.25in}
\fancyfootoffset[L]{2.25in}
\newcommand{\newaddition}[1]{\textcolor{black}{#1}}

\renewcommand{\emph}[1]{{\it #1}}

\begin{document}
\vspace*{0.2in}

\begin{flushleft}
{\Large
\textbf\newline{Statistical properties of 3D cell geometry from 2D slices} % Please use "sentence case" for title and headings (capitalize only the first word in a title (or heading), the first word in a subtitle (or subheading), and any proper nouns).
}
\newline
\\
Tristan A. Sharp\textsuperscript{1*},
Matthias Merkel\textsuperscript{2},
M. Lisa Manning\textsuperscript{2,3},
Andrea J. Liu\textsuperscript{1}
\\
\bigskip
\textbf{1} Dept. of Physics and Astronomy, University of Pennsylvania, Philadelphia, PA, USA
\\
\textbf{2} Physics Department, Syracuse University, Syracuse, NY, USA
\\
\textbf{3} Syracuse Biomaterials Institute, Syracuse, NY, USA
\\
\bigskip

% Use the asterisk to denote corresponding authorship and provide email address in note below.
* tsharp@sas.upenn.edu 

\end{flushleft}
\section*{Abstract}

Although cell shape can reflect the mechanical and biochemical properties of the cell and its environment, quantification of 3D cell shapes within 3D tissues \newaddition{remains} difficult, typically requiring digital reconstruction from a stack of 2D images.
We investigate a simple alternative technique \newaddition{to extract information about the 3D shapes of cells in a tissue; this technique connects the ensemble of 3D shapes in the tissue with the distribution of 2D shapes observed in independent 2D slices.
Using cell vertex model geometries, we find that the distribution of 2D shapes allows clear determination of the mean value of a 3D shape index.
We analyze the errors that may arise in practice in the estimation of the mean} 
3D shape index from 2D imagery and find that typically only a few dozen cells in 2D imagery are required to reduce uncertainty below 2\%.
This framework could be naturally extended to estimate additional 3D geometric features and quantify their uncertainty in other materials.

% Please keep the Author Summary between 150 and 200 words
% Use first person. PLOS ONE authors please skip this step. 
% Author Summary not valid for PLOS ONE submissions.   
% \section*{Author summary}

%\linenumbers

% Use "Eq" instead of "Equation" for equation citations.
\section*{Introduction}

Over the past decade, improved live-imaging techniques including multi-photon confocal~\cite{diaspro2001confocal} and light sheet microscopy~\cite{keller2008reconstruction} have dramatically altered our ability to quantify tissue architecture in \textit{in vivo} and \textit{in vitro} biological systems. In tandem, there has been an increased focus on developing mathematical models that can help organize and drive hypotheses about these complex systems.
 
Quite a bit of analysis and modeling has focused on confluent monolayers, where there are no gaps or overlaps between cells.  These two-dimensional sheets of tissue are often studied in cell culture systems~\cite{angelini2011glass,nnetu2013slow,park2015unjamming} and can also be found during embryonic development~\cite{chiou2012mechanical,farhadifar2007influence}. Much of that work focuses on understanding how cellular properties (interfacial tensions, adhesion, adherens junctions) give rise to local cellular shapes and also how they help to generate the large-scale, emergent mechanical properties of tissue.  
 
For example, researchers have developed a suite of mechanical inference techniques to estimate interfacial tensions and pressures from detailed images of cell shapes~\cite{brodland2014cellfit, chiou2012mechanical, YangMarchetti2017}.  Others have quantified precisely the deformation mechanisms in the developing fruit fly using dynamical shape changes~\cite{etournay2015interplay}. These methods rely heavily on automated watershed algorithms to segment membrane-labeled cell images in order to identify cell-cell interfaces in a network of many cells~\cite{farrell2017segga, mashburn2012enabling, soquet1998segmentation, fernandez2011oscillatory, krzic2012multiview, Etournay2016a}. Existing segmentation algorithms have largely been optimized to work on two-dimensional cell sheets.
  
Another set of experiments and models has focused on the statistics of cell shapes as a metric to quantify global mechanical tissue properties.  Specifically, studies of 2D cell vertex models (VMs) have found that cell shape may determine mechanical properties of confluent tissues (tissues with no gaps between cells)~\cite{bi2015density,bi2016motility, sussmanmerkel2017arxiv}. The models predict that when cells have a compact shape, so that their cross-sectional perimeter is small relative to their cross-sectional area, the tissue as a whole is solid-like in the sense that cells cannot migrate.  In contrast, when cells have an elongated shape, so that their perimeter is large relative to their area, then the tissue is \newaddition{fluid-like} in the sense that cells can easily exchange neighbors and migrate.  The transition from \newaddition{solid-like} to \newaddition{fluid-like} behavior is predicted to occur \newaddition{at} a specific value of the dimensionless 2D shape index, $p^{2D}$, which is defined as the ratio of the perimeter to the square root of the area. This prediction was shown to be precisely realized in human epithelial lung cell culture~\cite{park2015unjamming}.
  
Given that many biological tissues are fully three dimensional, it is natural to wonder whether any of this work can be extended to 3D. 
From a modeling perspective, it is straightforward, although technically challenging, to develop 3D simulations.  We have recently developed a 3D vertex-like model, called the 3D Voronoi model, and demonstrated that it, too, has a fluid-to-solid transition governed by cell shape \cite{merkel20173DVoronoi}. In this case, the governing shape parameter is $p^{3D}$, which is the dimensionless ratio of the surface area of each cell $S$ to its volume $V$: $p^{3D} = S/V^{2/3}$.  It also appears fairly straightforward to generalize mechanical inference methods to 3D \cite{Veldhuis2017a}.

Although advances in imaging techniques have allowed much clearer and deeper imaging of 3D structures, it remains a very open and technically difficult challenge to reconstruct the full network of cellular contacts in 3D~\cite{Khan4D, STEGMAIER2016225, 7533140, biotechniques2017}.  For example, watershed algorithms for segmentation will fail if there is even one 2D slice where the membrane structure is poorly resolved, and so in general they have a very large error rate in 3D.  In addition, many 3D structures of interest lie deep inside tissues where optical scattering makes live-imaging techniques difficult. In some cases, such as histological sections for staging of cancer tumors, only individual 2D images are available. Finally, a 3D reconstruction requires that all of the cells must remain stationary while an image stack is acquired.  Therefore, to our knowledge, very little of the exciting work in 2D can be robustly extended into live-imaged 3D experimental data.
   
This suggests that there may be an unexplored opportunity to use \emph{statistics} of 2D images, which are standard in the field, to infer something about the \emph{statistics} of 3D structures, an idea which has been exploited previously in materials science.  \newaddition{Methods to estimate the grain size distribution within poly-crystalline materials have been proposed that use processed 2D imagery and assume 3D grain shapes \cite{Takayama,Matsuura,militzer1999analysis}. Statistical} reconstruction of 3D structure from 2D imagery has \newaddition{also} been investigated for porous two-phase random media \cite{PhysRevE.58.224}, particulate media \cite{TALUKDAR2002419}, and media with shaped inclusions \cite{PhysRevLett.91.215506}. Typically, these methods start with a random 3D structure and have a process for evolving that structure to reduce differences between its 2D projections and 2D experimental data. 
   
In our case, we would like to understand whether we can infer useful 3D \newaddition{shape information from} 2D slices. Such an approach will not be directly helpful for mechanical inference methods, which rely on precise reconstructions of angles between junctions in 3D.  However, it could prove very useful for testing predictions of vertex-like models where tissue mechanics is predicted to depend on cell shape, or perhaps for testing models for \newaddition{studying constrained cell migration through complex networks. Such migration can lead to DNA damage that depends sensitively on the shapes and sizes of pores in the constraining environment~\cite{bennett2017elastic}.}

Therefore, the goal of this manuscript is to test whether information about 3D cell shapes can be reconstructed from randomly selected 2D image slices.  
\newaddition{Many experiments on mechanics and migration of cells in 3D focus on prepared tissues in collagen matrix or in centrifuged cell aggregates, and on other tissues, including organoids, certain tumors, and certain embryonic tissues, which appear isotropic and have relatively simple structure.
We therefore} perform this analysis in the context of a 3D Voronoi model, which is perhaps the simplest model for confluent cell bodies in 3D. In contrast to previous 3D inference methods, which typically sample only a fraction of the model phase space using Monte Carlo or dynamic minimization techniques\cite{PhysRevE.58.224, TALUKDAR2002419, PhysRevLett.91.215506}, the simplicity of this model allows us to quantify the relationship between 2D and 3D data across the entire relevant phase space.
   
We focus on determining whether the \newaddition{mean} 3D shape index, which models suggest is strongly correlated with tissue mechanics, can be inferred from 2D shape index, although we also explore other 2D and 3D shape descriptors in the Supplemental Material.  We find that there is a robust correlation between the 2D and 3D shape, and quantify the sensitivity of this correlation to sample size, experimentally relevant systematic errors, and tissue heterogeneity.  We find that relatively few cells are required to converge on the correct \newaddition{mean} 3D cell shape index, and that \newaddition{the estimates are quite robust with respect to moderate} errors in 2D cell perimeter measurements, dropping cells with a small cross-sectional area, \newaddition{and cell size heterogeneity}. This general framework may be extended to other contexts, ranging from more complicated \newaddition{or anisotropic} tissues to extracellular matrix to disordered inorganic materials; all that is required is to substitute the 3D Voronoi model with reasonable 3D models of such systems and to replace $p^{3D}$ with the relevant 3D shape descriptors.

% "Place figure captions after the first paragraph in which they are cited."
\begin{figure}[!h]
% The following line is meant to be removed before submission
%\center\includegraphics[width=0.5\linewidth]{SlicesOfVoroExLabel.png}
\center\includegraphics[width=\linewidth]{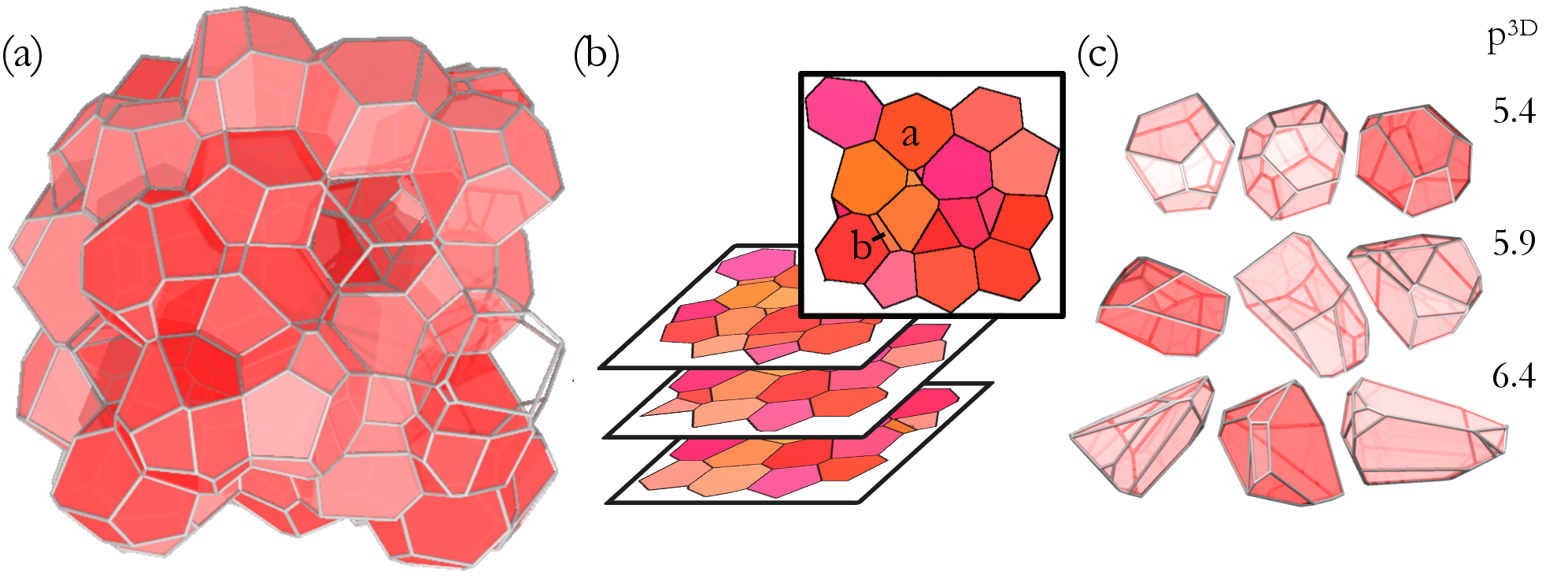}
\vspace{0.5cm}
\caption{{\bf \newaddition{Visualization of the simulated tissue, confocal cross-sections, and individual 3D cell shapes.}}
\newaddition{(a) A 3D tissue geometry in which all 3D cells have the same shape index, $p^{3D} = 5.40$, from the 3D Voronoi model.}
\newaddition{(b) The geometry seen in simulated ideal confocal imagery of that tissue.}
\newaddition{(c) Samples of the cell shapes from the model with specified $p^{3D}$ value.}
}
\label{fig1}
\end{figure}

% PLoS Template says: Results and Discussion can be combined.
\section*{Results}
\subsection*{2D cell shapes in slices of 3D cell packings}
As described in the methods section, we use the 3D Voronoi model to create cell packings with a specified, homogeneous 3D shape index, given by $p^{3D} = S/V^{2/3}$.
We \newaddition{cut the packings to obtain parallel slices of randomly oriented systems, yielding} 2D networks of edges and vertices, as illustrated in Fig~\ref{fig1}. For simplicity, we assume the slices to be extremely thin, although we can also vary the thickness of the optical section as described later. \newaddition{Our systems are small compared to experimentally-obtainable slices; for a large slice of an isotropic tissue, the results would be the same if one analyzed many different cells from the same slice.}  
We then calculate the 2D shape index $p^{2D}$ for each cell in the 2D slices.   

\begin{figure}[!h]
% The following line is meant to be removed before submission
\center\includegraphics[width=0.5\linewidth]{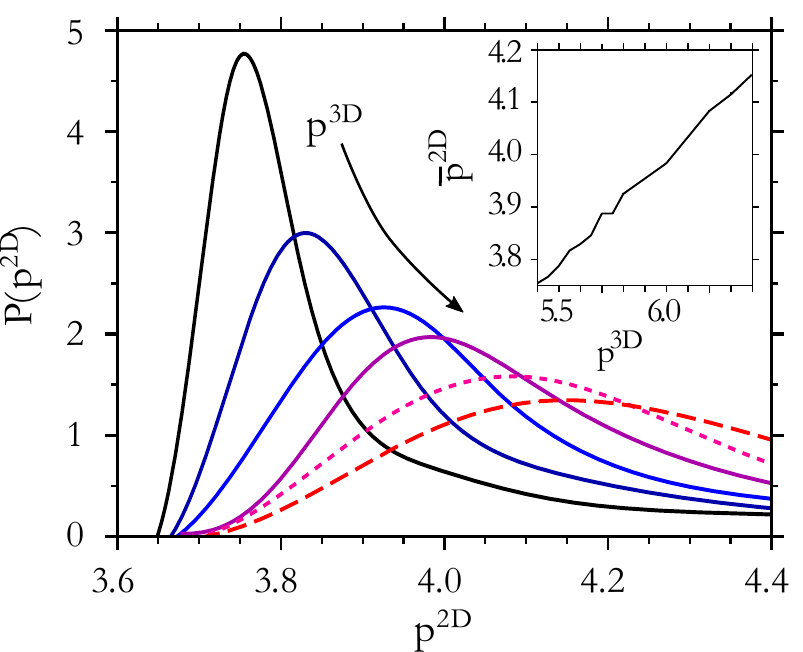}
\vspace{0.25cm}
\caption{{\bf 2D cell shape index distributions.}
The distribution of 2D shape index, $p^{2D}$, in slices from 3D cell packings provide a signature of the 3D shape index, $p^{3D}$.
$p^{3D}$ values vary from 5.4 (solid black) to 6.4 (dashed red) in increments of 0.2.
(Data available online\cite{repository}.) (Inset) The location of the distribution peak varies smoothly with $p^{3D}$.
}
\label{fig2}
\end{figure}

The distributions of $p^{2D}$ values for $20,000$ cells extracted from 500 3D packings are shown in Fig~\ref{fig2}, for several different values of $p^{3D}$.  Each curve demonstrates that even though each of the 3D cells has an identical 3D shape index, $p^{2D}$ exhibits a broad distribution of values.  The lowest possible value of $p^{2D}$ is that of a circle, $p^{2D}_{circ} = 2 \sqrt{\pi} \approx 3.54$, although compact shapes in tessellated patterns are typically higher. Fig~\ref{fig1} shows a compact cell (labeled ``a'') with a shape index of $p^{2D} = 3.75$. In contrast, very elongated shapes occasionally arise when polyhedral edges are nearly parallel to the slicing plane, as shown by the cell labeled ``b'' in Fig~\ref{fig1}.  These very elongated shapes are a purely geometric effect that contain little information about the underlying 3D shape index, as shown by \nameref{FigS1} in the SI.  They contribute to a long tail in the distribution of $p^{2D}$ that strongly affects both the mean and variance.  Therefore, we choose to focus on quantities such as the location of the peak of the distribution, $\overline{p}^{2D}$, and the half-width at half max (HWHM) that are less sensitive to these tails.

A first observation is that the peak in the 2D shape index ($\overline{p}^{2D}$) shifts dramatically with $p^{3D}$, suggesting that it should be possible to infer the 3D shape index from 2D data.  This is quantified by the inset to Fig~\ref{fig2}, which indicates that $\overline{p}^{2D}$ scales linearly with $p^{3D}$. Measures of sensitivity discussed in the SI confirm that the $p^{2D}$-distribution \newaddition{sensitively reflects} the $p^{3D}$ value. 

While vertex models suggest $p^{3D}$ is an important parameter governing tissue rheology, our approach can also be used to infer quantities that may be important for other models.  For example, some models for tissue mechanics are based on spherical particles, and in such systems the sphere volume fraction is an important parameter~\cite{fily2014freezing}.  Therefore, we also study a different 3D cell shape descriptor, $f_R$, the volume fraction of the largest sphere that can be inscribed within the cell.  As shown in \nameref{FigS2} in the SI, we find that \newaddition{the distribution of $p^{2D}$ also varies strongly with $f_R$, with a similar sensitivity, meaning that $f_R$ can be inferred as easily.}

Similarly, one might wonder whether there are better 2D shape descriptors than $p^{2D}$.  In the Supplemental Information we define another 2D observable that might be useful for quantifying cell shapes and tissue rheology, the anisotropy index $m$~\cite{Czajkowski2017}, which is the difference in the eigenvalues of the 2D shape tensor. Supplemental \nameref{FigS3} shows that $m$-distributions are also quite good (though not significantly better) at distinguishing between $p^{3D}$ or $f_R$ values.  Therefore in the following sections we focus on the shape indices in 2D and 3D ($p^{2D}$ and  $p^{3D}$), since they are simple to calculate and they control tissue rigidity in 2D and 3D cell vertex models.

%Other 3D shape descriptors, instead of $p^{3D}$, leave their signature in the distributions of 2D shape descriptors.

\subsection*{Precision of $p^{3D}$ estimation depends on sample size}

We ask how our estimate of $p^{3D}$ depends on the sample size (the number of $p^{2D}$ values extracted from cells in the 2D imagery, $N$).  In experiments on 2D lung epithelia\cite{park2015unjamming}, the mean shape index of cells only varies about 4\% in total. Therefore, the uncertainty in $p^{3D}$ must be smaller than a few percent. 

It can be arduous to obtain the $p^{2D}$ values upon which the $p^{3D}$ estimate is based. In some biology experiments, cell boundaries are digitized by an analyst who hand-traces the outlines of labeled membrane proteins on a computer.
Alternatively, automated computer algorithms may segment a 2D image into cell (and non-cell) regions, with trade-offs between speed and accuracy, to record the 2D shapes\cite{krzic2012multiview,fernandez2011oscillatory,mashburn2012enabling}. It is therefore important to establish how many cells must be analyzed to achieve the necessary level of accuracy in the estimate of $p^{3D}$.

To study how the best estimate of $p^{3D}$ and its uncertainty depends on the sample size, for each fixed value of $p^{3D}$ we segment a very large number of 2D cell images ($30,000$ cells) and generate a fixed reference distribution for $p^{2D}$. We repeat this process for $20$ values of $p^{3D}$ spaced equally between $5.4$ and $6.4$, to generate a library of reference distributions. These reference distributions are publicly available \cite{repository}.

\begin{figure}[!h]
% The following line is meant to be removed before submission
\center\includegraphics[width=0.5\linewidth]{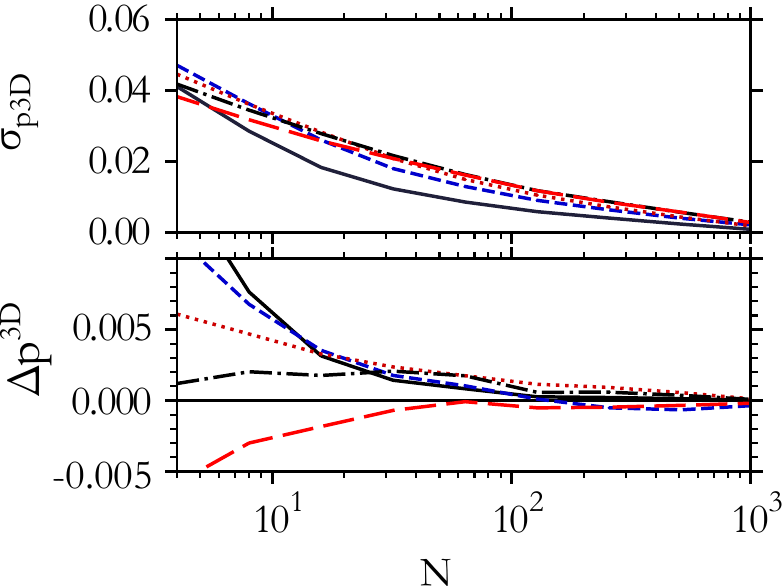}
\caption{{\bf \newaddition{Required number of measured cell shapes.}}
 The random error, $\sigma_{p^{3D}}$ (upper), and the systematic error, $\Delta p^{3D}$ (lower), in the $p^{3D}$ estimate both fall rapidly with the number of cells traced, $N$.  
The systematic error is much smaller than the random error.
The total error can be reduced to 1\% by \newaddition{calculating the shape index of} about 100 cells in this idealized case where the 2D measurements are exact.
Curves are for $p^{3D} = 5.5$ \newaddition{(black solid), 5.8 (blue dashed), 6.0 (red dotted), 6.1 (black dash-dotted), 6.2 (red long dashed)}. 
}
\label{convergence}
\end{figure}

Next, for a fixed value of $p^{3D}$ we segment $N$ randomly selected cells in 2D slices and generate a histogram of $p^{2D}$.
This histogram can be compared to the reference library. The standard Kolmogorov-Smirnov (K-S) test (described in the SI, Sec A.2), identifies the most likely distribution that produced the histogram, and correspondingly the estimated $p^{3D}$ value, $p^{3D}_{est}$. \newaddition{For this purpose we also created a publicly available online tool to compare data to the reference  distributions\cite{repository}}.
We repeat this process 1000 times to measure the spread of $p^{3D}_{est}$ estimates that occur.
The fractional random error of the $p^{3D}$ estimate is $\sigma_{p^{3D}} = \sigma_{p^{3D}_{est}}  / p^{3D}$, where $\sigma_{p^{3D}_{est}}$ is the standard deviation of $p^{3D}_{est}$.
The fractional systematic error is $\Delta p^{3D} = \langle p^{3D}_{est} - p^{3D} \rangle / p^{3D}$.
Fig~\ref{convergence} shows the random and systematic error as a function of number of cells traced.
Note the difference of y-axis scale; $\Delta p^{3D}$ is usually insignificant compared to $\sigma_{p^{3D}}$.
After tracing 50 cells at random, the random error in the $p^{3D}$ estimate is less than 2\%.
The estimate converges to the actual $p^{3D}$ with 1\% uncertainty after 100 random samples for $p^{3D}=5.4$ and after 200 random samples for $p^{3D}=5.8$. 
\newaddition{These results show that one may infer a sufficiently accurate estimate of $p^{3D}$ after imaging only a moderate number of cells.}

\subsection*{Estimation of $p^{3D}$ is sensitive to systematic tracing errors}
\label{ExperimentalError}

For the analysis in the previous section, we assumed that the 2D measurements of cell shape were exact, but data from experiments will have additional sources of noise and error. Therefore, we identify several likely sources and assess their impact on 3D shape estimation.

\newaddition{Cell shapes in 2D imagery can be measured by manual tracing on a computer or by automatic image segmentation and analysis.}
Some \newaddition{programs that measure cell shape} report the perimeter length as the number of pixels that the cell perimeter passes through, and this artificially raises the length of a line segment by an amount that depends on the angle relative to the pixel axes.
\newaddition{Attempting to infer the shape of} isolated cells rather than a compact cluster of cells that tessellates space can also artificially inflate cell perimeters.
Imprecise tracing of shape may result from other factors including inconsistent fluorescent dye saturation, uncertainty in identifying the cell borders, or limited image resolution, for example.
We can model these sources of error by distorting the images before processing them and then estimating $p^{3D}$.

For certain types of noise, the measured distributions of $p^{2D}$ can be directly computed from those without noise. For example, a systematic overestimate of perimeter by a fraction $\Delta L$ has the effect of scaling the distributions of Fig~\ref{fig2}(a) from $P(p^{2D})$ to $P(p^{2D}/(1+\Delta L))/(1+\Delta L)$. A random mis-estimate of perimeter $L$ by a fraction $\sigma_L$ is modeled by multiplying the perimeter by a Gaussian random factor (with standard deviation $L \sigma_L$).  This convolves the distributions of Fig~\ref{fig2}(a) with a Gaussian kernel.

\begin{figure}[!h]
% The following line is meant to be removed before submission
\center\includegraphics[width=\linewidth]{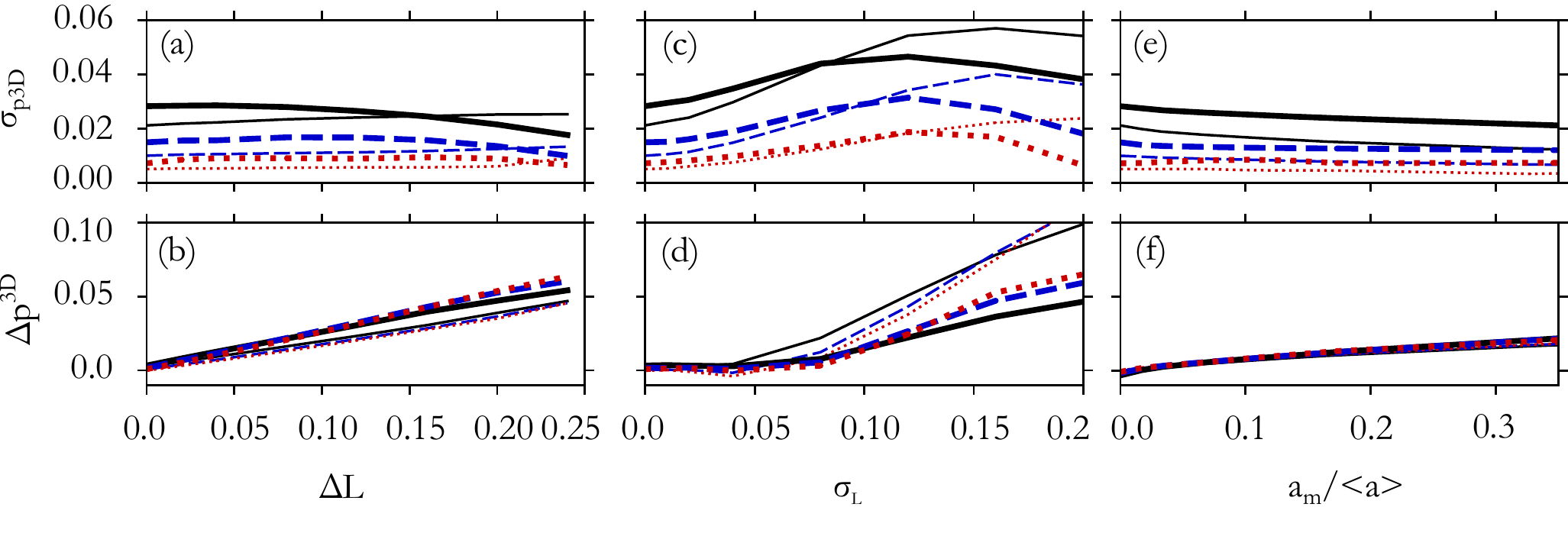}
\caption{{\bf Error propagation in estimate of $p^{3D}$.}
The fractional random error, $\sigma_{p^{3D}}$, and the systematic error, $\Delta p^{3D}$, in the $p^{3D}$ estimate increase with mis-measurement of cell perimeter, but are not significantly influenced by neglecting small cell areas.
Curves are for N = 16 (solid), 64 (dashed), and 256 (dotted) \newaddition{cells, and $p^{3D}=5.6$ (thick lines) and $p^{3D}=6.0$ (thin lines). There is little sensitivity to $p^{3D}$, so results for other $p^{3D}$ values are not shown.}
(a,b) A systematic overestimate of cell perimeter by a fraction $\Delta L$ produces a nearly-proportional systematic error in $p^{3D}$ estimate. Tracing additional cells reduces uncertainty but does not reduce systematic error.  
(c,d) 
A fractional random error of standard deviation $\sigma_L$
in the cell perimeter measurement also produces an error in the $p^{3D}$ estimate.
(e,f) Accidentally neglecting cells with small area causes only a small error.  
}
\label{errorrates}
\end{figure}

Using this information, we can, as in the previous section, compute the errors in the estimated $p^{3D}$, plotted in Fig~\ref{errorrates}.
Fig~\ref{errorrates}(a) shows that $\sigma_{p^{3D}}$, the random error in the $p^{3D}$ estimate, is nearly independent of the perimeter overestimation $\Delta L$ and decreases with the number of traced cells, to less than 1\% with N=256 cells.
$\Delta p^{3D}$, the systematic error, is approximately linear in $\Delta L$ and the error in the estimate can be reduced below 1\% if the perimeter error is less than 3\%.

Figs~\ref{errorrates}(c,d) show the error accrued from random tracing errors that result in mis-estimate of cell perimeter by a fraction $\sigma_L$.  
For large values $\sigma_L$, the estimated $p^{3D}$ can differ systematically from the true $p^{3D}$, but if $\sigma_L < 0.06$  (less than 6\% uncertainty of each $L$ measurement) the systematic error is less than 1\%, and tracing a large number of cells reduces the random error in $p^{3D}_{est}$.

Another possible source of systematic error is that the smallest shapes in microscopy may be accidentally overlooked and not traced. We generate reference distributions for this error by filtering the small-area shapes from the distribution functions. Neglecting shapes of area below a certain threshold, $a_m$, introduces relatively little error into the $p^{3D}$ estimate as shown in Fig~\ref{errorrates}.
When cells with area 10\% of the mean are accidentally neglected, the result is only an error of 1\% in $p^{3D}$. To avoid errors when comparing against the distributions generated here, cell shapes of all sizes should be included. If only large cell shapes can be traced reliably in experiment, then the resulting histograms may be compared to reference distributions that are generated using only the large shapes.

The above results show that generally an error in a tracing measurement generates a similar order of magnitude fractional error in the $p^{3D}$ estimate.
If multiple errors occur (\emph{e.g.} both $\Delta L$ and $\sigma_L$ are significant), and they are independent and uncorrelated, systematic errors $\Delta p^{3D}$ are approximately summed while random errors $\sigma_{p^{3D}}$ are approximately added in quadrature.

\subsection*{Assessing sensitivity to heterogeneity in $p^{3D}$}

So far, we considered homogeneous tissues where all cells have the same cell volume $V$ and 3D shape descriptor $p^{3D}$.  Of course, this is an idealization, as in real tissues there will be variations in these quantities. Here, we study the influence of such variations by generating 3D cell packings with Gaussian distributions of shapes $p^{3D}$ or volumes $V$. 

We first focus on variations in cell shape. We again generate the cell packings for an additional 200,000 cells as energy-minimized states of the 3D Voronoi model (see Methods), where states are included only if the cells achieve the target mean $\mu_{p^{3D}}$ and standard deviation $\sigma_{p^{3D}}$ of the 3D shape index; cases near extreme values of $\mu_{p^{3D}}$ (near 5.4 or 6.4)
targeting high-$\sigma_{p^{3D}}$  were not able to achieve the targets as seen in other simulations~\cite{merkel20173DVoronoi,sussmanmerkel2017arxiv} and so are excluded.

\begin{figure}[!h]
% The following line is meant to be removed before submission
\center\includegraphics[width=0.5\linewidth]{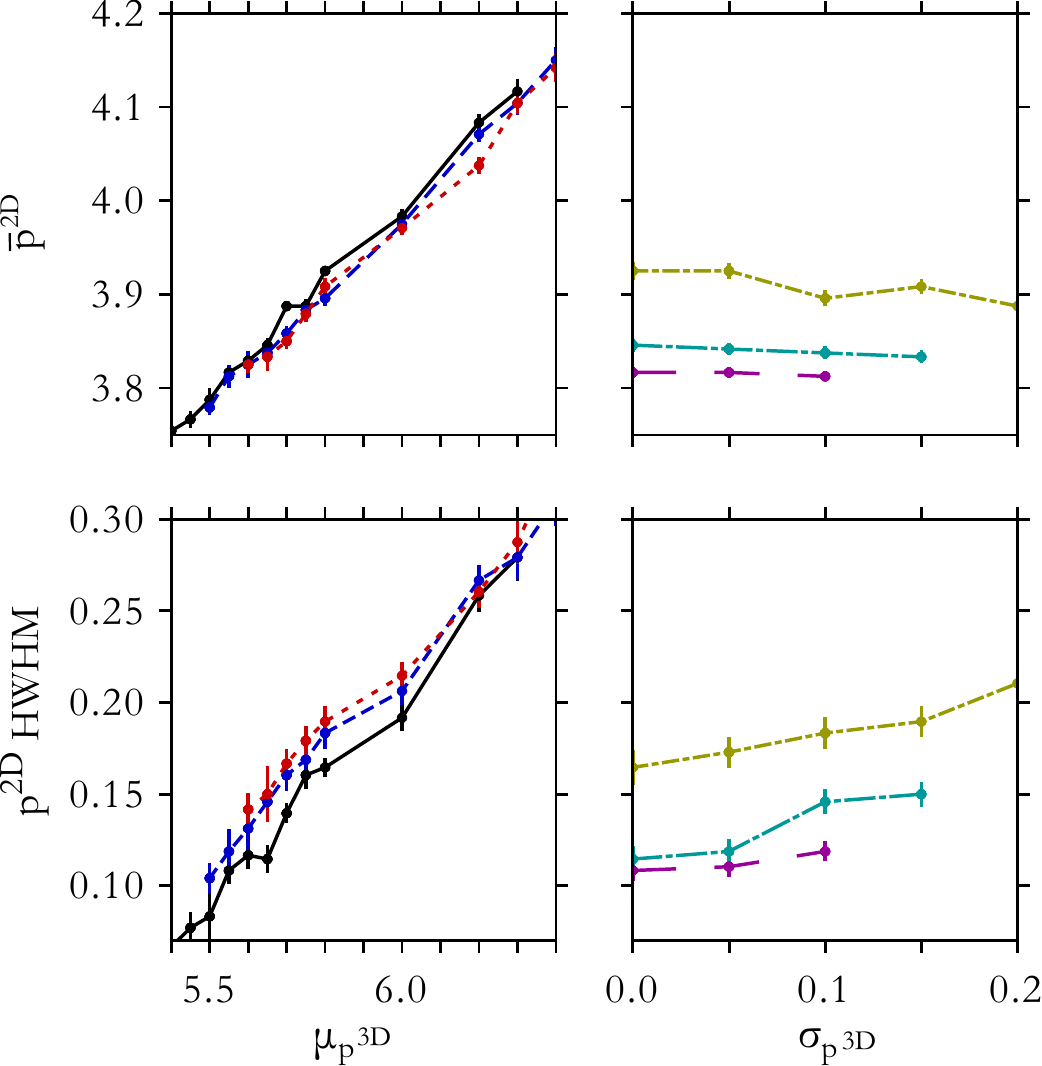}
\vspace{0.25cm}
\caption{{\bf Properties of $p^{2D}$ distributions from heterogeneous tissues.}
The peak and half-width-at-half-max (HWHM) of the $p^{2D}$-distribution accurately reflect the mean constituent 3D cell shape index, but are relatively insensitive to variation in shape. Curves vary $\mu_p^{3D}$ for $\sigma_{p^{3D}} = 0.0$ (black), 0.1 (blue, dashed), and 0.15 (red, dotted), and vary $\sigma_{p^{3D}}$ for $\mu_p^{3D} = 5.55$ (magenta, long dash), 5.65 (cyan, long dash-dot), and 5.8 (yellow, dash-dot).
}
\label{measure_by_cumulants}
\end{figure}

As discussed earlier, we focus on the location of the peak in the 2D shape index ($\overline{p}^{2D}$) and the half-width-at-half-max (HWHM) to minimize contributions of the universal tail.  
Fig~\ref{measure_by_cumulants} shows $\overline{p}^{2D}$ and HWHM for various values of $\mu_{p^{3D}}$ and $\sigma_{p^{3D}}$. While the peak and width of the distribution are strongly correlated with $\mu_{p^{3D}}$, these quantities are much less sensitive to $\sigma_{p^{3D}}$.
A sensitivity analysis based the K-S test confirms this result for both these and other shape descriptors, and similarly shows that fluctuations in volume have little impact on the estimations of cell shape (Supplemental Information Sec.~A.3 and \nameref{FigS4}).

Taken together, these results suggest that 2D shape analyses are very good at estimating the \emph{mean} 3D shape index, \newaddition{even with heterogeneity}, but are not very useful for estimating the variance of the 3D shape index. The lack of sensitivity is analyzed in Supplemental Sec. A.3, where we find that, although the $p^{2D}$-distributions vary with heterogeneity (and so the method is sensitive to heterogeneity in principle), the variation is \newaddition{comparatively} small. Future modeling work should therefore focus on understanding what features of heterogeneous systems are important for understanding global tissue mechanics.

\section*{Discussion}

We have investigated a method for determining statistical information about the 3D shapes of cells using only 2D slices through a tissue.
We focused on the 3D shape index $p^{3D}$ and found that it can be inferred directly and reliably from the distribution of measured $p^{2D}$ values.
We quantified the number of $p^{2D}$ measurements (and therefore cell traces) required to reduce the random and systematic errors in $p^{3D}$, showing that typically only of order 100 cell traces are required to reduce uncertainty in $p^{3D}$ below a given threshold.
The method is reasonably robust against the modeled sources of error and we suggest that the peak and width of the $p^{2D}$ distribution are easily-accessible quantities that biologists can compare directly to our reference distributions\cite{repository}.
We also study the effect of tissue heterogeneity, and find that 2D shapes can be used to estimate the average 3D shape, but provide little information about the variance in 3D shape.

It is possible to envision many extensions of this general framework. For example, confocal images occasionally have rather limited resolution along the z-axis, so that the effective thickness of the 2D slice is significant compared to the diameter of the cell. If the slice thickness is well-characterized, it is possible to use ray-tracing as discussed in the methods section to generate reference distributions relating 2D shapes to 3D shapes corresponding to a specific value of the slice thickness. Another extension would be to study different models for cellular structure, such as vertex models \cite{honda2004three,chiou2012mechanical,farhadifar2007influence} where the degrees of freedom are cell junctions instead of the cell centers, cellular Potts models \cite{chiang2016glass,graner1992simulation} where each cell is composed of a grid of points, or even models for non-biological materials such as foams~\cite{weaire2008pursuit}. 

\newaddition{We note that we are assuming that the tissue is isotropic on average.
For such tissues, the observed distribution of 2D shapes does not depend on the orientation of the 2D slice.  Conversely, for anisotropic tissues, a dependence on orientation will generally be observed, which could be used to infer the degree of alignment. We therefore presume that it is not an experimental challenge to determine whether a tissue is anisotropic. If it is not possible to study the dependence on orientation, then tissue anisotropy will show up in the 2D shape distributions, which will typically differ from those presented here.}

\newaddition{We further note that while it has been found that Voronoi tessellations approximate 2D tissue geometry reasonably well for many purposes\cite{Kaliman2015}, the degree to which the 3D Voronoi model mimics a specific 3D biological tissue can be expected to vary with the type of tissue.
On the other hand, because the high sensitivity of the $p^{2D}$ distribution to $p^{3D}$ originates from the fact that more compact cells produce more compact cross-sections, we expect the $p^{3D}$ estimate to be robust to minor variations in the geometry.}

It is also possible to envision extensions of this work beyond confluent cellular structures. For example, it should be possible to perform a similar analysis on particulate models and compare to nuclei-labeled images.  Alternatively, cell migration in fiber networks is conjectured to be limited by the rate at which cells can squeeze their nuclei through the pores in the mesh\cite{lichtman1970cellular}.  It would be interesting to see if one could estimate typical pore sizes from statistics of 2D slices of fiber network models.

Great progress is being made in developing techniques to fully reconstruct 3D cell shapes~\cite{Khan4D, STEGMAIER2016225, 7533140, biotechniques2017}, but their use is restricted to specimens that are optically transparent, or are compatible with light-sheet microscopy. Moreover, such techniques cannot be used in situations where cells exchange neighbors faster than a 3D scan can be completed.  In contrast, 2D images of cell structure are ubiquitous in medicine and biology, from histological sections of cancer tumors for use by pathologists to standard brightfield microscopy techniques, and they are also fast to obtain.

Our results suggest that the techniques and reference distributions developed here could be easily used by biologists and clinicians to obtain information about 3D shapes in the many cases where full 3D reconstruction is not possible.  In addition, this paper highlights a different way of thinking about how to use 2D cell images. Currently, cell shape and shape polarization are often described in terms of simple 2D quantities such as the axes of the best-fit ellipse or the length and width, and this naturally leads to a focus on how those quantities correlate with motion in a 2D plane. For cells in 3D tissue, our work suggests that 2D quantities are providing information about 3D shape, which could be used to drive and test hypotheses about cell migration \newaddition{and tissue mechanics} in 3D.

\section*{Methods}

\newaddition{To investigate the connection between 3D cell shape and 2D cross-sectional cell shapes in a 3D geometry that is representative of simple isotropic 3D tissues, we use} the recently developed 3D Voronoi model \cite{merkel20173DVoronoi}.
Based on a Voronoi tessellation of cell center positions~\cite{rycroft2009voro}, we divide a simulation volume with periodic boundaries into polyhedra, representing cells.
Cell centers move to minimize a Hamiltonian,
\begin{eqnarray*}
\label{eq:cellmodel}
  E =  k_S \sum_{i=1}^N (S_i - S_{0i})^2 +  k_V \sum_{i=1}^N (V_i - V_{0i})^2 \text{,}
\end{eqnarray*}
% "Equations should be referred to as: We see in Eq~(\ref{eq:schemeP}) something." 
in which each cell $i$ has a target volume $V_{0i}$ and surface area $S_{0i}$, and  $N$ is the number of cells.
We choose the target cell parameters either to be equal for all cells ($S_{0i} = S_0$ and $V_{0i} = V_0$) to simulate a homogeneous system, or such that target volumes $V_{0i}$ and target shape indices $p_{0i}^{3D}=S_{0i}/V_{0i}^{2/3}$ are independently drawn from Gaussian distributions.
The parameters $k_S$ and $k_V$ are stiffness constants which we set to $1$ here without loss of generality, since we will focus on cases in the fluid regime of the model~\cite{merkel20173DVoronoi} where all cells attain their target parameters and $E \approx 0$.
The simulation volume is fixed at $V_{total}=\sum_i V_{0i}$.
Starting from uniformly randomly distributed cell centers, we use the FIRE minimizer\cite{PhysRevLett.97.170201} to minimize the energy with respect to all cell positions, generating an isotropic ensemble of 3D cell shapes.

For homogeneous target shapes, we observe that when $p_{0}^{3D} \gtrsim 5.41$, all cells satisfy their optimal values, $S_i = S_0$ and $V_i = V_0$.
This allows generation of a large ensemble of disordered cells all with the same shape index, $p^{3D} = S/V^{2/3}$.
When $p_{0} = S_0/V_0^{2/3} \lesssim 5.4$, the geometry becomes pinned~\cite{merkel20173DVoronoi} such that the mean shape index is $\langle p^{3D} \rangle \approx 5.4$, because it appears that disordered packings with smaller shape index do not exist\cite{merkel20173DVoronoi}.
For values of $p_{0} > 6.4$, multifold vertices become common and minimization algorithms face challenges~\cite{sussmanmerkel2017arxiv} and so we restrict simulations to $p_{0} \leq 6.4$.
We created packings with shape indices $p^{3D}$ from 5.4 to 6.4 with 0.05 increments.

Once 3D packings with a defined distribution of $p^{3D}$ are generated, we simulate the acquisition of 2D cross sections by generating images of intersections of cells with a specified plane (Fig~\ref{fig1}) using the software POV-Ray~\cite{povray}.
Based on a segmentation of these cellular cross sections, we quantify the cell 2D shape index $p^{2D}$ as the quotient of perimeter divided by the square root of the cross-sectional area, which provides us with a histogram of 2D shape indices $p^{2D}$ for the given ensemble of cell packings with its predefined $p^{3D}$ distribution.

To compare different $p^{2D}$ distributions with each other, we use two kinds of measures.  As a practical measure for experimentalists to extract the 3D shape index from a $p^{2D}$ distribution, we propose to use peak and HWHM of the $p^{2D}$ distribution (Figs. 1 and 5).
As a complementary measure to compare $p^{2D}$ distributions in more detail, we also use the K-S test, which measures the maximum distance between the two respective cumulative distributions \cite{eadie1971statistical} (used for Figs. 2-4 and described in detail in the Supplemental Information).

Extension of the method to other 2D and 3D cell geometry descriptors is discussed in the Supplemental Information.

\section*{Supporting information}

% Include only the SI item label in the paragraph heading. Use the \nameref{label} command to cite SI items in the text. [They say filename should match label. -TAS]
\paragraph*{Attachment 1.}
\label{Supplemental}
{\bf Supplemental Information} 
The attached document supports the supplemental figures.

\paragraph*{Fig S1}
\label{FigS1}
% The following line is meant to be removed before submission
\includegraphics[width=0.5\linewidth]{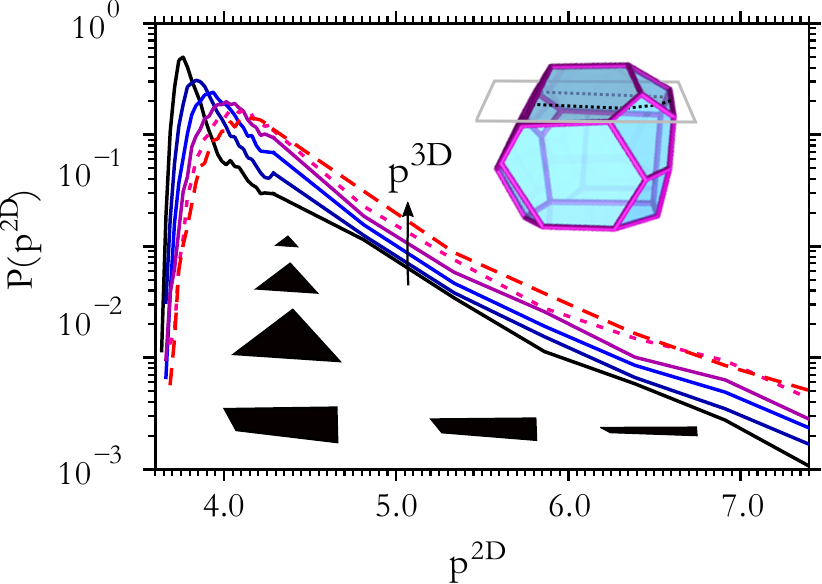} \\
{\bf The tail of the 2D cell shape index distribution.} The $p^{2D}$-distribution tail decays with a similar form for all values of $p^{3D}$ indicating that little information about $p^{3D}$ is contained in the tail. Lines correspond to $p^{3D}$ values from 5.4 (solid black) to 6.4 (dashed red), varying in increments of 0.2.
Example cross sections illustrate that arbitrarily large $p^{2D}$ values can be produced even from small-$p^{3D}$ (\textit{i.e.} compact) polyhedra, when the cross section is nearly parallel to a nearby edge---this produces nearly-rectangular slim shapes, for which $p^{2D}$ diverges when the cutting plane is near the edge. In contrast, general cross sections near a vertex are triangles, and the shape index, which is insensitive to the size of the triangle, retains a constant constant value independent of proximity of the cutting plane and vertex.

\paragraph{Fig S2}
\label{FigS2}
% The following line is meant to be removed before submission
\includegraphics[width=0.5\linewidth]{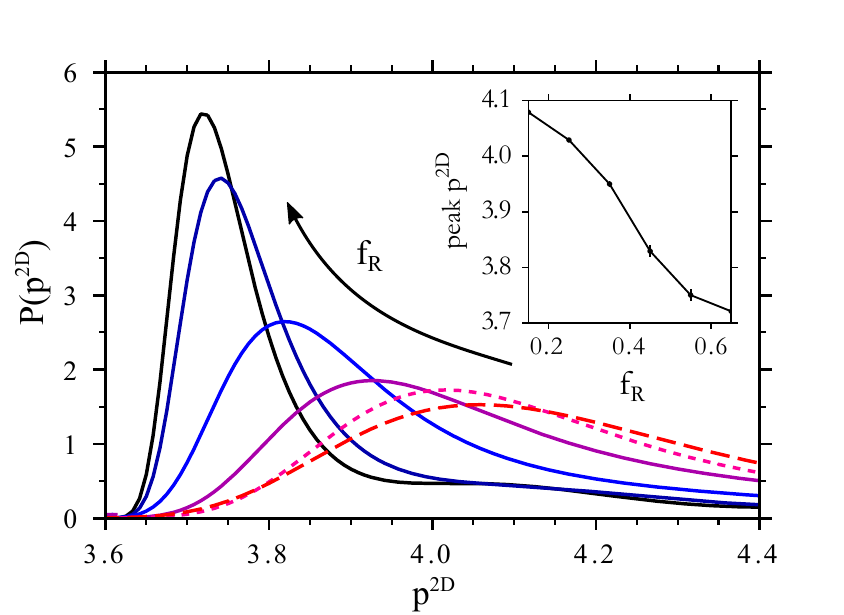} \\
{\bf The distribution of $p^{2D}$ values depends on the volume fraction of the largest inscribed sphere $f_R$, discussed in Sec.~A.1.}  The volume fraction of the largest inscribed sphere, $f_R$, discussed in the Supplemental Information, Sec.~A.1, can be estimated from the $p^{2D}$-distribution. Shown are $p^{2D}$ histograms, created from cell packings created with the same range of $p_0^{3D}$ as in the main text, and binning the cells with respect to their $f_R$. Each curve corresponds to a different $f_R$ bin, where $f_R$ values vary in 0.1 increments from 0.15 (dashed, red) to 0.65 (solid, black).

\paragraph*{Fig S3}
\label{FigS3}
% The following line is meant to be removed before submission
\includegraphics[width=0.5\linewidth]{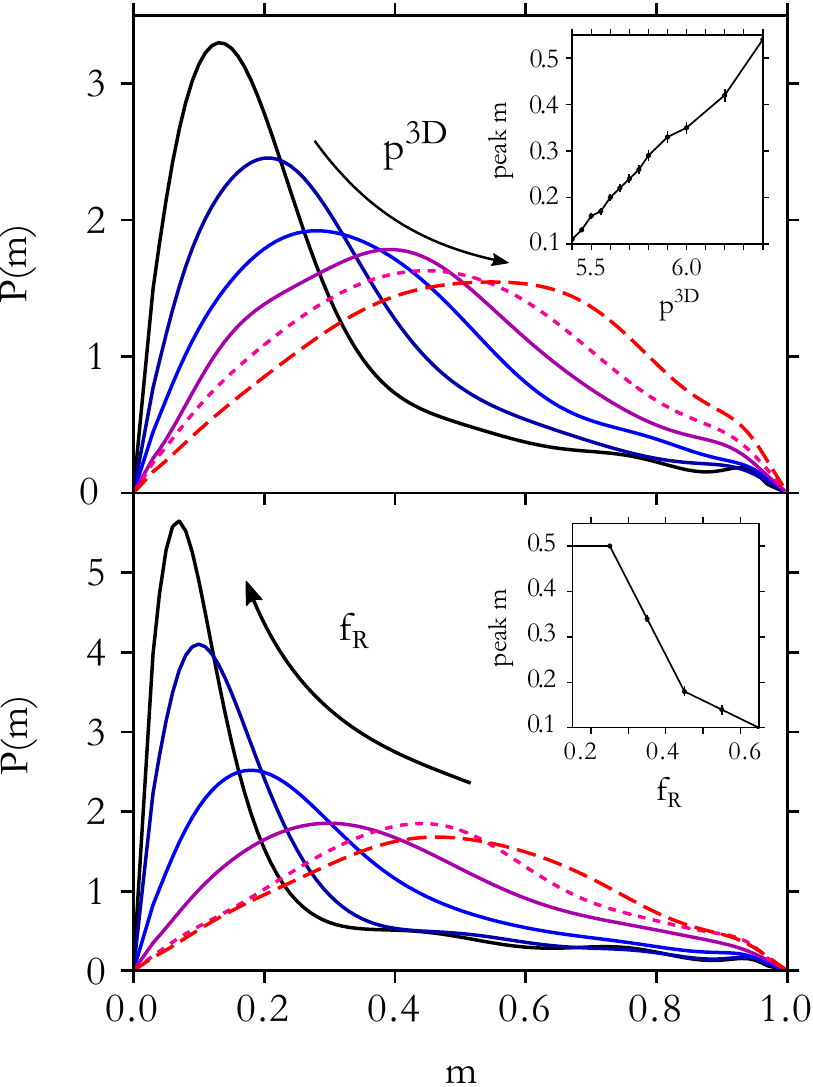} \\
{\bf $m$-distribution depends on 3D cell shape index $p^{3D}$ and volume fraction of the largest inscribed sphere $f_R$, discussed in Sec.~A.1.}  Distributions of the 2D anisotropy index $m$ reflect the 3D shape descriptors $p^{3D}$ and $f_R$ as discussed in Sec A.1. As in previous figures, $p^{3D}$ varies from 5.4 (solid black) to 6.4 (dashed red) in increments of 0.2.
$f_R$ values vary in 0.1 increments from 0.15 (dashed, red) to 0.65 (solid, black).

\paragraph*{Fig S4}
\label{FigS4}
% The following line is meant to be removed before submission
\includegraphics[width=\linewidth]{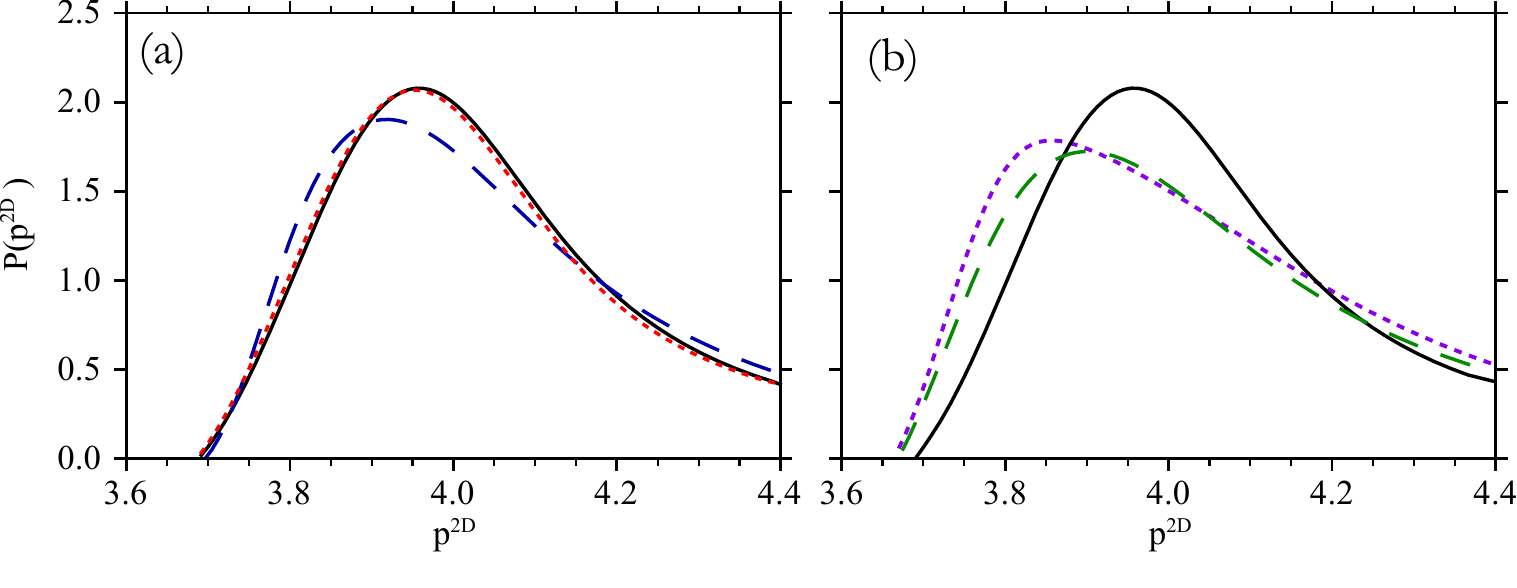} \\ 
{\bf $p^{2D}$-distributions for heterogeneous 3D cell shape, discussed in Sec.~A.3.} 
(a) The $p^{2D}$-distribution for cell packings of homogeneous 3D shape index $p^{3D}=5.9$ (solid black) and of heterogeneity in 3D shape index $\sigma_{p^{3D}} = 0.15$ (dashed blue) and in cell volume $\sigma_V = 0.15 V_0$ (dotted red).
(b)
The $p^{2D}$-distribution for cell packings of homogeneous 3D shape index $p^{3D}=5.9$ (solid black) is compared to a packing of a binary mixture (dashed green). The binary mixture is 50\% of $p^{3D} = 5.6$ and 50\% of $p^{3D} = 6.2$ cells, 
so that the mean 3D shape index is $\langle p^{3D} \rangle = 5.9$.
The superposition (dotted violet) of the $p^{2D}$-distributions of packings with homogeneous $p^{3D} = 5.6$ and $p^{3D} = 6.2$, respectively, are a close approximation of the heterogeneous system.

\section*{Acknowledgments}

We thank Paul Janmey and Anne van Oosten for suggesting this investigation and for helpful discussions.

%This project was supported by the National Cancer Institute of the National
%Institutes of Health under Physical Sciences Oncology Center (PSOC)
%award No. U54 CA193417 (TAS and AJL). We also acknowledge the Simons Collaboration on Cracking the Glass problem (454947) for initiating this collaboration. MM and MLM acknowledge funding from the Alfred P.\ Sloan Foundation, the Gordon and Betty Moore Foundation, the Research Corporation for Scientific Advancement, and computational support through NSF ACI-1541396. In addition, we gratefully acknowledge support from the National Science
%Foundation (NSF) under NSF-DMR-1506625 (TAS) and NSF-DMR-1352184 and NSF-PHY-1607416 (MLM), as well as from the Simons Foundation (445222 to MLM and 327939 to AJL).

%\nolinenumbers

%\bibliography{confocal}

% Either type in your references using
% \begin{thebibliography}{}
% Text
% \end{thebibliography}
%
% or
%
% Compile your BiBTeX database using our plos2015.bst
% style file and paste the contents of your .bbl file
% here. See http://journals.plos.org/plosone/s/latex for 

%% SUPPLEMENTAL
\newpage
\pagebreak
\setcounter{page}{16}
% TAS - I think Title should be copy-pasted from main text.
\begin{flushleft}
	{\Large
		\textbf\newline{Supplemental information: \\ Statistical properties of 3D cell geometry from 2D slices} 
	}
	\newline
	\\
	Tristan A. Sharp\textsuperscript{1},
	Matthias Merkel\textsuperscript{2},
	M. Lisa Manning\textsuperscript{2,3},
	Andrea J. Liu\textsuperscript{1}
	\\
	\bigskip
	\textbf{1} Dept. of Physics and Astronomy, University of Pennsylvania, Philadelphia, PA, USA
	\\
	\textbf{2} Physics Department, Syracuse University, Syracuse, NY, USA
	\\
	\textbf{3} Syracuse Biomaterials Institute, Syracuse, NY, USA
	\\
	\bigskip
\end{flushleft}

\renewcommand{\thesection}{A}

\subsection{Alternative geometry descriptors}

The main text focuses on cell shape indices $p^{2D}$ and $p^{3D}$ as 2D and 3D shape descriptors, respectively.
However, the method is quite general and other shape descriptors can also be considered.  Here we demonstrate this point with two other quantities. First we show that the $p^{2D}$ distribution can be used to extract the mean value of $f_R$, the ratio of the max inscribed sphere volume to Voronoi cell volume, which provides another characterization of a 3D shape. To study the distribution of $p^{2D}$ values, we use the same set of simulations of 20,000 cells as in the main text and bin cells by their $f_R$ value. Fig~S2 shows that the distribution of $p^{2D}$ varies with $f_R$ value, which ranges between 0.1 and 0.7.
The distributions in Fig~S2 resemble those in Fig~2, because both $p^{3D}$ and $f_R$ characterize compactness of the 3D shapes.

We additionally ask if a different 2D shape descriptor can be used to extract the mean value of $p^{3D}$. We consider distributions of $m$, the anisotropy index\cite{yang2017correlating}. 
The shape anisotropy, $m$, is determined using the $2\times 2$ moment of inertia tensor of the shape,
\begin{equation*}
G = \int_{area} \hspace{2mm} \vec{v} \otimes \vec{v} \hspace{1mm} dx \hspace{1mm} dy .
\end{equation*}
Here, the integral runs over the area of the shape in the x-y plane.
$\vec{v}(x,y)$ is the 2D vector from the shape center of mass.
The anisotropy index, $m$, is the difference of the eigenvalues of the matrix $G$ divided by their sum.
As discussed in the main text, a confocal slice through a collection of cells of a specified $p^{3D}$ produces a distribution of these 2D shape descriptors.  

The additional distributions of $m$, for varied values of $p^{3D}$ and $f_R$, are shown in Fig~S3, in analogue to the distributions of Fig~2 in the main text. As $m$ remains bounded within the unit interval, the distribution lacks a long tail and may be preferable in some cases to $p^{2D}$ as a 2D shape  descriptor. The presence of the peak at low-$m$ is reflective of the correlations between $m$ and $p^{2D}$. 

Multi-dimensional (joint) distributions $P_{p^{3D}}(p^{2D},m,a)$ may also be used for more complete comparisons between the 3D model and the 2D geometry in slices. Additional 2D descriptors provide more sensitivity and points of comparison between experiment and model, whereas additional 3D descriptors provide more information about the 3D geometry.

\pagebreak

\subsection{K-S test for goodness of fit and sensitivity}
\label{sect:ks}

The Kolmogorov-Smirnov (K-S) test characterizes the likelihood of a set of measurements, given a probability distribution for the measured quantity.\cite{eadie1971statistical}
For our purposes, K-S assigns a distance, $D$, to the dissimilarity of a measured histogram and one of the reference distribution, $P(p^{2D})$.  
Specifically, $D$ is computed from the maximum separation between the empirical cumulative distribution function (EDF) of the data and the cumulative distribution function (CDF) of the reference distribution.
The CDF of a distribution $P(X)$ is $C(x) = \int_{-\infty}^{x} dx' P(x')$.
The corresponding quantity for the set of $p^{2D}$ values, $\left\{ X_i \right\}$, that go into the histogram is the EDF, $C_{\text{emp}}(x) = \frac{1}{N} \sum_{i=1}^N \Theta(x - X_i)$. $N$~is the number of samples in the histogram, and 
$\Theta(x - X_i)$ is the Heaviside step function, equal to 0.0 for $x$ less than $X_i$ and otherwise equal to 1.0.
The K-S distance $D$ is the maximum separation between the two, $D = \text{max}_{x} \hspace{0.5mm} | C_{\text{emp}}(x) - C(x) |$.  This distance allows identification of the model distribution that best fits the sampled data in the main text.

The K-S test quantifies the goodness of fit between data and model in the presence of random noise\cite{eadie1971statistical}. This is done by comparing the K-S statistic $\sqrt{N} D$ with a published table and indicates in our context whether the 2D geometry from the experiment is consistent with the 3D geometry of the model considered.

Furthermore, the K-S distance $D$ can be used to quantify the sensitivity of a distribution to $p^{3D}$ by characterizing how quickly a $p^{2D}$-distribution changes upon changing $p^{3D}$. For a small change of $p^{3D}$ by $\delta$, we find that the K-S distance $D$ increases from zero linearly. The sensitivity is given by $d D / d \delta$. The $p^{2D}$-distributions of Fig~2 are most sensitive to $p^{3D}$ at $p^{3D} \approx 5.4$ where $d D / d \delta \approx 1.5$, while near 6.4 the distributions are less than half as sensitive with $d D / d \delta \approx 0.5$. Thus, while the location of the distribution peak $\overline{p}^{2D}$ varies linearly with $p^{3D}$ with slope 0.4, the changes in \emph{distribution shape} produce additional sensitivity near $p^{3D} = 5.4$.

Similarly, the sensitivity of the $m$-distribution (Sec.~A.1) to $p^{3D}$ is found to be $d D_{p^{3D},m} / d \delta \approx 1.4$ at $p^{3D} = 5.4$ and falls to $d D_{p^{3D},m} / d \delta \approx 0.4$ by $p^{3D} = 6.4$. 
Thus distributions of $m$ and $p^{2D}$ are about equally sensitive to $p^{3D}$, and the uncertainties in Fig 3 and Fig 4 in the main text would be similar for $m$. 
Here $D_{p^{3D},m}$ is the K-S distance between $m$-distributions from $p^{3D}$ values that differ by $\delta$. Analogously, $D_{f_R,p^{2D}}$ being the K-S distance between $p^{2D}$-distributions from $f_R$ values that differ by $\delta_{f_R}$, 
the sensitivity to $f_R$ is found to be $d D_{f_R,p^{2D}} / d \delta_{f_R} \approx 0.5$ and $d D_{f_R,m} / d \delta_{f_R} \approx 0.6$ near $f_R = 0.2$, rising to $d D_{f_R,p^{2D}} / d \delta_{f_R} \approx 2.7$ and $d D_{f_R,m} / d \delta_{f_R} \approx 2.3$ at $f_R = 0.5$. This quantifies the level of sensitivity that is evident in Fig S2 and Fig S3.

\subsection{Analysis of fluctuations in heterogeneous systems}

In the main text, we focused on systems in which the preferred 3D shape index is the same for all cells. However, real tissues may be heterogeneous with different preferred shape indices for different cells.  We first show that 2D shapes are relatively insensitive to volume heterogeneity. Specifically, we simulated the vertex model for cells with the same preferred shape index for each cell but with a  
Gaussian distribution of target cell volumes (mean $\mu_V$ and standard deviation $\sigma_{V}$).
Supplemental Fig~S4(a) shows distributions for the homogeneous case, $\sigma_{V}=0$ (solid black), and a heterogeneous case with $\sigma_{V}=0.15 \mu_V$ (dotted red). Likewise, we study a system in which the preferred surface area is the same for all cells but the preferred volume varies such that the 3D shape index, $p^{3D}$ is drawn from a Gaussian distribution with the same mean value and with standard deviation $\sigma_{p^{3D}}$ (dashed blue curve in Fig~S4(a)). The curve corresponding to heterogeneous volume is much more similar to the homogeneous case than the one corresponding to a similarly heterogeneous distribution of cell shapes ($\sigma_{p^{3D}}=0.15$, also plotted in Fig~S4(a), dashed blue). The K-S distance discussed in Sec.~A.2 quantifies this; the distance between the dotted red curve and the black curve is $d D/ d \sigma_{p^{3D}} \approx 0.1$, which is about 10\% of the distance between the black curve and the dashed blue curve. Practically speaking, this means that the 2D shape data presented in this manuscript Figs~2,~5,~S2,~and~S3 provide accurate estimates of 3D cell shape even if there are fluctuations in preferred cell volume.

It is also natural to ask \emph{why and how} the 2D shapes are more sensitive to mean 3D shape.  Specifically, for heterogeneous systems with a distribution of 3D shapes, both the peak and the width of the 2D shape distribution was sensitive to the mean 3D shape, but they were not very sensitive to the variance in 3D shape or volume. 

To gain an intuition for this observation, it is useful to look at a specific example. Fig~S4(b) illustrates the 2D shape distribution for a binary mixture of approximately 50\% $p^{3D}=5.6$ and 50\% $p^{3D}=6.2$ cells.
The distribution $P$ from the mixed system (Fig~S4(a) dashed green line) is distinct from $P$ of the homogeneous systems, the most similar of which is $p^{3D}=5.9$ (solid black line).
This emphasizes that the 2D shape distribution from a mixed system is distinct from the distributions of homogeneous systems.

Next, we plotted the superposition of the homogeneous distributions, $P_{mix} \approx 0.5 P_{5.45} + 0.5 P_{6.0}$ (Fig~S4(b), violet dotted line) and found good agreement with the distribution of the mixed system.
This suggests that the mixed system is composed of approximately the same 3D shapes that are generated in homogeneous simulations of the constituent $p^{3D}$ values.
Under this approximation, the $p^{2D}$-distribution of the heterogeneous system is
\begin{equation}
\label{admixture}
P_{mix} = \sum_\beta c_\beta f_\beta P_{p_\beta^{3D}} 
\end{equation}
where $f_\beta$ is the fraction of cells of that are in subpopulation $\beta$, and the correction factor $c_\beta \approx 1.0$.
The correction factor $c_\beta$ for subpopulation $\beta$ accounts for the increased number of intersections of the imaging plane with large of elongated 3D cells than with small or compact 3D cells.
It is related to the mean 2D area of cells of that subpopulation, $\langle a \rangle_\beta$, and is given by $c_\beta = \sum_\beta f_\beta \langle a \rangle_\beta / \langle a \rangle_\beta$
where the sum occurs over each constituent cell population $\beta$ with a specific size $V$ and shape $p^{3D}$.

Since the 2D shape distribution from heterogeneous systems is approximately given by the superposition of pure-$p^{3D}$ distributions, the sensitivity to $p^{3D}$-heterogeneity is linked to the variation with $p^{3D}$ itself (\textit{e.g.} Fig~1).
The lack of change of the distribution with heterogeneity simply arises from two facts. First, many similar constituent distributions underlie the distribution of the heterogeneous system, so for instance, Gaussian-distributed $p^{3D}$ heterogeneity with $\sigma_{p^{3D}}=0.15$, 68\% of the distribution weight comes from the narrow range $\mu_{p^{3D}} - 0.15 < p^{3D} < \mu_{p^{3D}} + 0.15$. 
Second, the smooth variation of the distribution with $p^{3D}$ (\textit{e.g.} Fig~1) means that Gaussian heterogeneity (which is symmetric above and below the mean $p^{3D}$) produces somewhat canceling differences to the homogeneous distribution.

%\nolinenumbers
%\bibliography{confocal}

% Either type in your references using
% \begin{thebibliography}{}
% Text
% \end{thebibliography}
%
% or
%
% Compile your BiBTeX database using our plos2015.bst
% style file and paste the contents of your .bbl file
% here. See http://journals.plos.org/plosone/s/latex for 
% \begin{thebibliography}{10}
% \bibitem{bib1}
% Ohno S.
% \newblock Evolution by gene duplication.
% \newblock London: George Alien \& Unwin Ltd. Berlin, Heidelberg and New York:
%   Springer-Verlag.; 1970.

% \bibitem{bib2}
% Magwire MM, Bayer F, Webster CL, Cao C, Jiggins FM.
% \newblock {{S}uccessive increases in the resistance of {D}rosophila to viral
%   infection through a transposon insertion followed by a {D}uplication}.
% \newblock PLoS Genet. 2011 Oct;7(10):e1002337.

% \end{thebibliography}

\end{document}